\documentclass[12pt,a4paper]{article}
\pdfoutput=1
\usepackage{amsmath,amssymb,exscale}
\usepackage{amsfonts,amsthm}
\usepackage{graphicx, float, epsfig}
\usepackage{mathrsfs}
\usepackage{slashed}
\usepackage{textcomp}
\usepackage[T1]{fontenc}
\usepackage[utf8]{inputenc}
\usepackage[table]{xcolor}
\usepackage[lowtilde]{url}
\usepackage{cancel}
\usepackage[colorlinks=true,urlcolor=blue,anchorcolor=blue,citecolor=blue,filecolor=blue,linkcolor=blue,menucolor=blue,pagecolor=blue,linktocpage=true]{hyperref}
\numberwithin{equation}{section}
\usepackage{verbatim} 
\usepackage{appendix}
\usepackage{caption}
\usepackage{float}
\usepackage{tikz}
\usetikzlibrary{calc,shapes,arrows,patterns,positioning,decorations.markings,trees}
\tikzset{>=stealth'}
\tikzstyle point=[minimum size=1mm,inner sep=0pt,outer sep=0pt,shape=circle,fill=black]
\newcommand{\HIDDEN}[1]{}
\usepackage{ulem}
\definecolor{fillcolor}{HTML}{C4C4C4}
\definecolor{plotgray}{HTML}{626262}
\def\crbig{\\\noalign{\vspace {3mm}}}
\newcommand{\be}{\begin{equation}}
\newcommand{\ee}{\end{equation}}
\newcommand{\ba}{\begin{eqnarray}}
\newcommand{\ea}{\end{eqnarray}}
\newcommand{\bay}{\begin{array}{rcl}}
\newcommand{\eay}{\end{array}}
\newcommand{\ra}{\rightarrow}

\begin{document}
	
\title{Towards 6D Little String Theory of Particles}

\author{Risto Raitio\footnote{E-mail: risto.raitio@helsinki.fi}\\	
Helsinki Institute of Physics, P.O. Box 64, \\00014 University of Helsinki, Finland}

\date{February 16, 2024}  \maketitle 

\abstract{\noindent  
A model for particles based on preons in chiral, vector and tensor/graviton supermultiplets of unbroken global supersymmetry is engineered. The framework of the model is little string theory. Phenomenological predictions are discussed.
	
\vskip 1.5cm


\newpage
\tableofcontents
\vskip 1.0cm

\section{Introduction}
\label{intro}

We investigate the possibility that at high enough energy the elementary particles may not be the standard model (SM) particles. We have proposed instead a supersymmetric preon scenario for first two flavors of fermions. In addition, we introduce tensor multiplet fields, two extra dimensions, i.e. the little string theory (LST), and holographic duality.

What we see in nature depends, of course, on the resolving power of apparatus available, as well as on our preferred theoretical concepts. At laboratory energy scale we observe quarks and leptons as point like particles. The celebrated SM gauge principle has been successful with three interactions. At the (unreachable) string energy scale we would expect to "detect" superstrings. Consequently, we have proposed that the key symmetry should be supersymmetry rather than an internal symmetry, not forgetting the latter though. Due to lack of experimental evidence for supersymmetry on quark and lepton level, we have introduced supersymmetric preons,\footnote{~ Chernons have no direct experimental support either.} called here chernons, which occur between the superstring scale and the inflationary reheating scale $T_R$ in the early universe.

Chernons are free particles above the energy scale $\Lambda_{cr}$, numerically about $\sim 10^{10} - 10^{16}$ GeV. It is close to reheating scale $T_R$ and the grand unified theory (GUT) scale. We conjecture chernons obey superconformal field theory (SCFT). At $\Lambda_{cr}$ preons make a phase transition by an attractive Chern-Simons (CS) model interaction into composite states of standard model quarks and leptons, including gauge interactions. 

This note is organized as follows. In section \ref{kinetic} we extend our two flavor chernon model to include color of the SM. Section \ref{potential} recaps the binding  chernon-chernon interaction. The symmetries of the schematic little string theory model are discussed in subsection \ref{lst} and tensormultiplets in subsection \ref{tensormultiplets}. Holographic duality is introduced subsection \ref{holodual} to define approximate gravity. Conclusions are given in section \ref{conclusions}. Appendix \ref{basicidea} is provided to visualize the difference between the standard model supersymmetry and ours. 

This note contains copiously review material to make it self-contained. Crucial new material to our scenario is added in sections \ref{kinetic} and \ref{6dmodels}. The type of this note is exploratory phenomenology: to search for ideas and concepts which would substantiate the existence of a consistent model beyond the standard model. The present scenario should be considered as a first step modeling.

\section{Extending the Wess-Zumino action}
\label{kinetic}

The divisive point of the chernon model for visible and dark matter is the following: we think it is dubious to add to each known SM particle its (unobserved) superpartner. Instead, supersymmetry should be implemented so that all particles needed to describe nature are written together with their superpartners in  the Lagrangian ((\ref{chiralwz}) - (\ref{chiralcolor})) of the model. Our method was introduced in \cite{rai_00, rai_01}. The result turned out to have close resemblance to the Wess-Zumino (WZ) model \cite{Wess_Z}, which contains three neutral fields: a Majorana spinor $m$, the real fields $s$ and $p$ with $J^P = \frac{1}{2}^+, 0^+$, and $0^-$, respectively. The kinetic WZ Lagrangian is
\be
\mathcal{L}_{\rm{WZ}} = -\frac{1}{2} \bar{m}\cancel\partial m - \frac{1}{2} (\partial s)^2 - \frac{1}{2} (\partial p)^2
\label{chiralwz}
\ee
where $m$ and $s$ form the chiral multiplet. We assume that the pseudoscalar $p$ is the axion \cite{Peccei_Q}, and denote it below as $a$. It has a fermionic superparther, the axino $n$, a candidate for dark matter but not discussed further here.

To include charged matter we define the following charged chiral field Lagrangian for fermion $m^-$, complex scalar $s^-$ and the electromagnetic field tensor $F_{\mu\nu}$
\be
\mathcal{L}_{\rm{WZ}_{Charge}} = -\frac{1}{2}m^- \cancel\partial m^- - \frac{1}{2} (\partial s^-)^2 - \frac{1}{4}F_{\mu\nu}F^{\mu\nu}
\label{chiralcharge}
\ee

We set color to the neutral fermion $m \ra m^0_i$ ($i = R, G, B$) in (\ref{chiralwz}). The color sector Lagrangian is then
\be
\mathcal{L}_{\rm{WZ}_{Color}} = -\frac{1}{2}                                                                        \sum_{{i=R,G,B}}\Big[\bar{m}^0_i\cancel\partial m^0_i - \frac{1}{2} (\partial g_i)^2 \Big]
\label{chiralcolor}
\ee

We now have the supermultiplets shown in table \ref{tab:table1}.

\begin{table}[H]
	\begin{center}
		\captionsetup{width=.8\linewidth}
		\begin{tabular}{|l|l|} 
			\hline
			Multiplet & Particle, Sparticle \\ 
			\hline   
			chiral multiplets spins 0, 1/2 & 
			$s^-$, $m^-$;~a, n \\
			vector multiplets spins 1/2, 1 & $m^0$, $\gamma$;~$m_i, g_i$ \\  
			\hline
		\end{tabular}
		\caption{\small The particle $s^-$ is a neutral scalar particle. The particles $m^-, m^0$ are charged and neutral, respectively, Weyl spinors. The a is axion and n axino. $m^0$ is color singlet particle and $\gamma$ is the photon. $m_i$ and $g_i$ (i = R, G, B) are zero charge color triplet fermions and bosons, respectively.}
		\label{tab:table1}
	\end{center}
\end{table}   

Note that in table \ref{tab:table1} there is a zero charge quark triplet $m_i$ but no gluon octet. Instead, supersymmetry demands the gluons to appear only in triplets at this stage of cosmological evolution. The dark sector we get from (\ref{chiralcolor}).} 

The matter-chernon correspondence for the first two flavors (r = 1, 2; i.e. the first generation) is indicated in table \ref{tab:table2}. 

\begin{table}[H]
	\begin{center}
		\captionsetup{width=.8\linewidth}
		\begin{tabular}{|l|l|} 
			\hline
			SM Matter 1st gen. & Chernon state \\ 
			\hline                                       
			$\nu_e$ & $m^0_R m^0_G m^0_B$ \\     
			$u_R$ & $m^+ m^+ m^0_R$ \\			  
			$u_G$ & $m^+ m^+ m^0_G$ \\  			
			$u_B$ & $m^+ m^+ m^0_B$ \\
			
			$e^-$ & $m^- m^-m^-$ \\		         
			$d_R$ & $m^- m^0_G m^0_B$ \\		 
			$d_G$ & $m^- m^0_B m^0_R$ \\			
			$d_B$ & $m^- m^0_R m^0_G$ \\
			\hline
			W-Z Dark Matter & Particle \\
			\hline   
			boson (or BC) & $s$, axion(s) \\
			$e'$ & axino $n$ \\
			meson, baryon $o$ & $n\bar{n}, 3n$ \\
			nuclei (atoms with $\gamma ')$ & multi $n$ \\
			celestial bodies & any dark stuff \\	 
			black holes & anything (neutral) \\
			\hline
		\end{tabular}
		\caption{\small Visible and Dark Matter with corresponding particles and chernon composites. $m^0_i$ (i = R, G, B) is color triplet, $m^{\pm}$ are color singlets of charge $\pm 1/3$. $e'$ and $\gamma'$ refer to dark electron and dark photon, respectively. BC stands for Bose condensate. Chernons obey anyon statistics.}
		\label{tab:table2}
	\end{center}
\end{table}

After quarks have been formed by the process described in section \ref{potential} the SM octet of gluons will emerge because it is known that fractional charge states have not been observed in nature. To make observable color neutral, integer charge states (baryons and mesons) possible we proceed as follows. The local $SU(3)_{color}$ octet structure is formed by quark-antiquark composite pairs as follows (with only color charge indicated):
\be
\rm{Gluons}: \footnotesize{R\bar{G}, R\bar{B}, G\bar{R}, G\bar{B}, B\bar{R}, B\bar{G}, \frac{1}{\sqrt{2}}(R\bar{R}-G\bar{G}), \frac{1}{\sqrt{6}}(R\bar{R}+G\bar{G}-2B\bar{B})}
\label{gluons} 
\ee

With the gluon triplet the first hunch is that they form, with octet gluons now available, the $3 \otimes 3 \otimes 3 =10 \oplus 8 \oplus 8 \oplus 1$ bosonic states with spins 1 and 3. These three gluon coupling states would need a separate investigation, see e.g. \cite{Papa_A_F}. 

Finally, we introduce the weak interaction. After the SM quarks, gluons and leptons have been formed at scale $\Lambda_{cr}$ there is no more observable supersymmetry in nature \cite{rai_05}. To avoid a more complicated vector supermultiplet in table \ref{tab:table1}, we may append the standard model electroweak interaction in our model as an empirical fact. The standard model has now been heuristically derived.

\section{Chernon-chernon interaction}
\label{potential}

This section is covered in \cite{rai_03, rai_03a} with references to original papers.\footnote{~We wish to add that the CS term can also be added to models which are not topological quantum field theories. In 3D, this gives rise to a massive photon.} The chernon-chernon scattering amplitude in the non-relativistic approximation is obtained by calculating the t-channel exchange diagrams of the Higgs scalar and the massive gauge field. The propagators of the two exchanged particles and the vertex factors are calculated from the action \cite{Beli_D_F_H}.

The gauge invariant effective potential for the scattering considered has been obtained in \cite{Kogan, Dobroliubov}
\begin{equation}
	V_{{\rm CS}}(r)=\frac{e^{2}}{2\pi }\left[ 1-\frac{\theta }{m_{e}}\right]
	K_{0}(\theta r)+\frac{1}{m_{e}r^{2}}\left\{ l-\frac{e^{2}}{2\pi \theta }%
	[1-\theta rK_{1}(\theta r)]\right\} ^{2} 
	\label{Vmcs}
\end{equation}
where $K_{0}(x)$ and $K_{1}(x)$ are the modified Bessel functions and $l$ is the angular momentum ($l=0$ in this note). In (\ref{Vmcs}) the first term $[~]$ corresponds to the electromagnetic potential, but it now behaves like a Yukawa potential, the second one $\{~\}^2$ contains the centrifugal barrier $\left(l/mr^{2}\right)$, the Aharonov-Bohm term and the two photon exchange term.

In (\ref{Vmcs}) the first term may be positive or negative while the second term is always positive. The function $K_{0}(x)$ diverges as $x \ra 0$ and approaches zero for $x \ra \infty$ and $K_{1}(x)$ has qualitatively similar behavior. For our scenario we need negative potential between all chernons, including equal charge ones. We must have the $K_{0}(\theta r)$ term dominating in (\ref{Vmcs}) with the condition\footnote{~For applications in condensed matter physics, one must require $\theta \ll m_{e}$, and the scattering potential given by (\ref{Vmcs}) then comes out positive \cite{Beli_D_F_H}.}
\be
\theta \gg m_e
\label{condition}
\ee
The potential (\ref{Vmcs}) also depends on $v^{2}$, the vacuum expectation value, and on $y$, the parameter that measures the coupling between fermions and Higgs scalar. Being a free parameter, $v^{2}$ indicates the energy scale of the spontaneous breakdown of the $U(1)$ local symmetry.

\section{Six dimensional theories}
\label{6dmodels}

The framework of section \ref{kinetic} is little string theory (the name was given in \cite{LMS}). It is obtained as an effective theory of parallel and overlapping NS 5-branes with 16 supercharges in the limit when gravity decouples in type IIA or type IIB string theory. The 6D LST (subsection \ref{lst}) is UV complete. The supersymmeric gravity (SUGRA) (subsection \ref{tensormultiplets}) is not UV complete with matter \cite{deser.stelle.kay:1977, ricc.sagn:1998}. Using holographic duality we can introduce approximate gravity to LST (in subsection \ref{holodual}).

\subsection{Little string theory}
\label{lst}

The string theory vision predicts the existence of new non-local theories of which we are interested in the case of D = 6. We begin by reviewing \cite{Vafaetal.2016}. See also the more recent paper \cite{DelZottoetal.2022, ahmedaetal.2023}. LSTs are generated by stacks of (Neveu–Schwarz) NS5-branes. These branes are decoupled from the bulk without taking the low energy limit $\alpha ' \ra 0$ \cite{aharony&al:2004}. At high energies, they can become tensionless and support new degrees freedom required for a UV completion of the theory.

The non-local ingredients of a theory of extended objects, such as strings, can be seized by a quantum field theory with a local stress energy tensor, while the string scale $M_{string}$ remains. The UV completion, however, is not a quantum field theory. The local characterization breaks down as we reach the string scale $M_{string}$.

The little string theories allow studying stringy extended objects but with fewer complications. In known constructions these theories exhibit properties which are typical of closed string theories with tension set by $M_{string}^2$. Here we approach LST from bottom up with F-theory \cite{Vafa.1996} in the background.

In F-theory there is a non-compact base $B$ of complex dimension two. This is supplemented by an elliptic fibration to reach a non-compact Calabi-Yau threefold. In the resolved phase, the intersection pairing of the base coincides with the Dirac pairing \cite{Argyresetal.2022} for two-form potentials of the theory on its tensor branch. For a 6D superconformal field theory, one demands that the Dirac pairing is negative definite. For an LST, it is required that this pairing is negative semidefinite. 

In field theory terms, this is usually enforced by the condition that all gauge theoretic anomalies are canceled on the tensor branch of the theory. In F-theory terms, classifying LSTs thus amounts to determining all possible elliptic Calabi-Yau threefolds which support a base $B$ with negative semidefinite intersection pairing. All LSTs are given by a small extension of 6D SCFTs \cite{Vafaetal.2016}.

Back to field theory terms, we find that the string charge lattice of any LST with more than one tensor multiplet is an affine extension of the string charge lattice of an SCFT, with the minimal imaginary root of the lattice corresponding to the little string charge. Hence, all LSTs arise from an affine extension of SCFTs.

An emblem of all known LSTs is T-duality. We disclose some of the conditions necessary to realize a supersymmetric little string theory. 6D supersymmetric theories should admit a tensor sector. Several LSTs include some dynamical tensor multiplets and vacua parameterized at low energies by vevs of scalars in these tensor multiplets. To reach a point of strong coupling one has to tune the vevs of the dynamical scalars to zero. In addition to dynamical tensor multiplets, one has to allow the possibility of non-dynamical tensor multiplets which set mass scales for the 6D supersymmetric theory.

In a theory with tensor multiplets, one has scalars $S^{I}$ and their bosonic superpartners $B^{-,I}_{\mu \nu}$, with anti-self-dual field strengths as indicated in table \ref{tab:table3}.

\begin{table}[H]
	\begin{center}
		\captionsetup{width=.8\linewidth}
		\begin{tabular}{|l|l|} 
			\hline
			Multiplet & Particle, Sparticle \\ 
			\hline
			tensormultiplet & $S^I$,  $B^{-,I}_{\mu\nu}$ \\
			\hline
		\end{tabular}
		\caption{\small The tensormultiplet.} 
	\label{tab:table3}
	\end{center}
\end{table}  

The vevs of the $S^{I}$ govern, among other things, the tension of the effective strings which couple to these two-form potentials. In a theory with gravity, one must also include an additional two-form potential $B_{\mu \nu}^{+}$ coming from the graviton multiplet. Given this collection of two-form potentials, we get a lattice of string charges $\Lambda_{string}$ \cite{seib-tayl.2011}, and a Dirac pairing
\begin{equation}
	\Lambda_{string} \times \Lambda_{string} \rightarrow \mathbb{Z}
\end{equation}
in which we allow for the possibility that there may be a null space for this pairing. It is convenient to describe the pairing in terms of a matrix $A$ in which all signs have been reversed. Thus, we can write the signature of $A$ as $(p,q,r)$ for $q$ self-dual field strengths, $p$ anti-self-dual field strengths, and $r$ the dimension of the null space.

For a 6D supergravity theory with $T$ tensor multiplets, the signature is $(T,1,0)$. Even more is true in a 6D theory of gravity. Diffeomorphism invariance enforces the condition that $\det A = -1$ \cite{seib-tayl.2011}.

In supersymmetric theories decoupled from gravity one arrives at the necessary condition that the signature of $A$ is $(p,0,r)$. In this case each of the two-form potentials has a real scalar superpartner denoted as $S^{I}$. The kinetic term for these scalars is
\begin{equation}
	\mathcal{L}_{eff} \supset A_{IJ} \partial S^{I} \partial S^{J}
\end{equation}

Observe that if $A$ has a zero eigenvector, some linear combinations of the scalars will have a trivial kinetic term. When this occurs, these tensor multiplets define parameters of the effective theory on the tensor branch, they are non-dynamical fields.

This leaves one with two general possibilities. Either $A$ is positive definite (i.e. $A > 0$), or it is positive semidefinite (i.e. $A \geq 0$). To reach a 6D SCFT, a necessary condition is $A > 0$ \cite{Bhardwaj:2015xxa}. We summarize the various possibilities for self-consistent 6D theories in table \ref{tab:table4}.

\begin{table}[H]
	\begin{center}
	\begin{tabular}
		[c]{|c|c|c|c|}\hline
		Theory & 6D SUGRA & 6D LST & 6D SCFT\\\hline
		signature & $(T,1,0)$ & $(p,0,r)$ & $(T,0,0)$\\\hline
		$\det A:$ & $ \det A  = - 1$ & $\det A=0$ & $\det
		A>0$\\\hline
	\end{tabular}
	\caption{\small Values of det A for 3 different 6D theories.}
	\label{tab:table4}
	\end{center}
\end{table}

We introduce LST as a 6D theory with $\det A = 0$.\footnote{~Moving down in energy, one flows down from SUGRA to LST and then eventually to SCFT.} Some linear combinations of the scalar fields for tensor multiplets will have trivial kinetic term, they are dimensionful parameters. In a 6D theory with a single gauge group factor and no dynamical tensor multiplets this parameter is just the overall value 
$S_{null} = 1 / g_{YM}^2$, with $g_{YM}$ the Yang-Mills 
coupling of a gauge theory. This YM theory contains solitonic solutions which one can identify with strings
\begin{equation}\label{soliton}
	F = - \ast_{4} F,
\end{equation}
meaning that we dualize in the four directions transverse to an effective string. One can expect $A$ to contain some general null space, and with each null direction, a non-dynamical tensor multiplet of parameters
\begin{equation}
	\overrightarrow{v}_{null} \equiv N_{1} \overrightarrow{v}^{1} + ... + N_{T} \overrightarrow{v}^{T} \,\,\,\text{such that} \,\,\, A \cdot \overrightarrow{v}_{null} = 0.
\end{equation}
for the two-form potential, and
\begin{equation}
	S_{null} = N_{1} S^{1} + ... + N_{T}S^{T}
\end{equation}
for the corresponding linear combination of scalars. Since they specify dimensionful parameters, we get an associated mass scale, which is referred to as $M_{string}$
\begin{equation}
	S_{null} = M_{string}^2.
\end{equation}

Returning to our example from 6D gauge theory, the tension of the solitonic string in equation (\ref{soliton}) is just $1/g_{YM}^2 = M_{string}^2$. At energies above $M_{string}$, the effective field theory is no longer valid, and one must provide a UV completion.

On general grounds, $A \geq 0$ could have many null directions. However, in the case where we have a single interacting theory, that is $A$ is simple, there are further strong restrictions. 
When $A \geq 0$ is simple, all of its minors are positive definite: $A_{minor} > 0$. Consequently, there is precisely one zero eigenvalue, and the eigenvector is a positive linear combination of basis vectors. There is only one dimensionful parameter $M_{string}$. This also means that if one deletes any tensor multiplet, one reaches a positive definite intersection pairing and a 6D SCFT. What we have just learned is that if we work in the subspace orthogonal to the ray swept out by $S_{null}$, then the remaining scalars can all be collapsed to the origin of moduli space. When one does this, one reaches the LST limit. This property of the matrix $A$ is referred to as the tensor-decoupling criterion for an LST. The fact that decoupling any tensor multiplet takes one to an SCFT imposes sharp restrictions.

This discussion has up to now focused on some necessary conditions to reach a UV complete theory different from a 6D SCFT. In \cite{Bhardwaj:2015xxa, Seiberg:1996qx} the specific case of 6D supersymmetric gauge theories was considered, and closely related consistency conditions for UV completing to an LST were presented. One sees the same consistency condition $A \geq 0$ appearing for any effective theory with (possibly non-dynamical) tensor multiplets.

Though we have given a number of necessary conditions that any putative LST must satisfy, to truly demonstrate their existence we must pass beyond effective field theory, embedding these theories in a UV complete framework such as string theory. One therefore has to turn to the F-theory realization of little string theories. F-theory provides a formulation which systematically enumerates possible tensor branches. This is discussed in \cite{Vafaetal.2016, DelZottoetal.2022}, but it is beyond the scope of this brief phenomenological note. Instead, we review holographic models in subsection \ref{holodual}.

\subsection{Supergravity tensor multiplets}
\label{tensormultiplets}

We now review briefly minimal $(1,0)$ 6D supergravity coupled to $n$ tensor multiplets. Supersymmetry in 6D is generated by an $Sp(2)$ doublet of chiral spinorial charges $Q^a$ $(a=1,2)$, obeying the symplectic Majorana condition
\be 
Q^a = \epsilon^{ab} C \bar{Q}^T_b
\ee 
where $\epsilon^{ab}$ is the $Sp(2)$  antisymmetric invariant tensor. Since all  fermion fields appear as $Sp(2)$ doublets, we use $\Psi$ to denote a doublet $\Psi^a$.  

The  theory in \cite{romans} includes the vielbein $e_{\mu}^a$, a left-handed gravitino $\Psi_{\mu}$, $(n+1)$  antisymmetric tensors $B^r_{\mu\nu}$ $(r=0,...,n)$  obeying (anti)self-duality conditions, $n$ right-handed tensorini $\chi^m$  $(m=1,...,n)$, and $n$ scalars. The scalars parameterize the coset space $SO(1,n)/SO(n)$, and are  thus associated to the $SO(1,n)$ matrix $(r=0,..., n)$
\begin{center}
	V = 	
	$\begin{pmatrix}
		v_r \\ x^m_r 
	\end{pmatrix}$ 
\end{center}

\begin{table}[H]
	\begin{center}
		\captionsetup{width=.8\linewidth}
		\begin{tabular}{|l|l|} 
			\hline
			Multiplet & Particle, Sparticle \\ 
			\hline
			tensormultiplet & $g_{\mu\nu}$, $\Psi_{\mu}$ \\
			 & $B^+_{\mu\nu}$, $\chi^m$ \\
			\hline
		\end{tabular}
		\caption{\small The tensormultiplet.} 
		\label{tab:table5}
	\end{center}
\end{table}
\noindent 
whose matrix elements satisfy the constraints
\ba
v^r v_r =& 1  \nonumber\\ 
v_r v_s - x^m_r x^m_s =& \eta_{rs} \nonumber \\
v^r x^m_r =& 0   
\label{scalars}
\ea
Defining
\be 
G_{rs} = v_r v_s + x^m_r x^m_s 
\ee 
the tensor (anti)self-duality conditions can be succinctly written
\be 
G_{rs} H^{s \mu\nu\rho} =\frac{1}{6e} \epsilon^{\mu\nu\rho\alpha\beta\gamma} H_{r \alpha\beta\gamma}\quad ,
\label{selfdual}
\ee 
where $H^r_{\mu\nu\rho}= 3 \partial_{[\mu} B^r_{\nu\rho ]}$. These relations only hold to lowest order in the fermion fields, and imply that  $v_r H^r_{\mu\nu\rho}$ is self dual, while the $n$ tensors $x^m_r H^r_{\mu\nu\rho}$ are antiself dual, as one can see using (\ref{scalars}). The divergence of (\ref{selfdual}) yields the second-order tensor equation
\be 
D_{\mu} (G_{rs} H^{\sigma\mu\nu\rho} )=0  
\label{tensoreq}
\ee 
while, to lowest order, the fermionic equations are
\be
\gamma^{\mu\nu\rho} D_{\nu} \Psi_{\rho} + v_r H^{r\mu\nu\rho} \gamma_{\nu} \Psi_{\rho} -
\frac{i}{2} x^m_r H^{r\mu\nu\rho} \gamma_{\nu\rho} \chi^m + 
\frac{i}{2} x^m_r \partial_{\nu} v^r \gamma^{\nu} \gamma^{\mu} \chi^m = 0 
\ee 
and
\be
\gamma^{\mu} D_{\mu} \chi^m -\frac{1}{12} v_r H^{r\mu\nu\rho} \gamma_{\mu\nu\rho} \chi^m - \frac{i}{2} x^m_r
H^{r\mu\nu\rho} \gamma_{\mu\nu} \Psi_{rho} -
\frac{i}{2} x^m_r \partial_{\nu} v^r \gamma^{\mu} \gamma^{\nu} \Psi_{\mu} =0 
\ee 
Varying the fermion fields in them with the supersymmetry transformations 
\ba
\delta e_{\mu}^a =& -i (\bar{\epsilon} \gamma^a \Psi_{\mu} ) \\ 
\delta B^r_{\mu\nu} =& i v^r (\bar{\Psi}_{[\mu} \gamma_{\nu]} \epsilon )
+\frac{1}{2} x^{mr} (\bar{\chi}^m \gamma_{\mu\nu} \epsilon )  \nonumber\\ 
\delta v_r =& x^m_r (\bar{\epsilon} \chi^m ) \nonumber\\
\delta \Psi_{\mu} =& D_{\mu} \epsilon +\frac{1}{4} v_r H^r_{\mu\nu\rho} \gamma^{\nu\rho} \epsilon 
\nonumber\\ \delta \chi^m =& \frac{i}{2} x^m_r
\partial_{\mu} v^r \gamma^{\mu} \epsilon + \frac{i}{12} x^m_r H^r_{\mu\nu\rho} \gamma^{\mu\nu\rho} \epsilon   \label{susy}
\ea 
generates the bosonic equations, using also eqs. (\ref{selfdual}) and (\ref{tensoreq}). Thus, the scalar field equation is
\be 
x^m_r D_{\mu}(\partial^{\mu} v^r ) +\frac{2}{3} x^m_r v_s H^r_{\alpha\beta\gamma}  H^{s \alpha\beta\gamma} =0 \quad 
\ee 
while the Einstein equation is
\be 
R_{\mu\nu} -\frac{1}{2} g_{\mu\nu} R + \partial_{\mu} v^r \partial_{\nu} v_r - \frac{1}{2} g_{\mu\nu} \partial_{\alpha} v^r \partial^{\alpha} v_r - G_{rs} H^r_{\mu\alpha\beta} H^{s~~\alpha\beta}_{\nu} = 0
\ee 
To this order, this amounts to a proof of  supersymmetry,  and it is also possible to show that the  commutator of two supersymmetry transformations on the bosonic fields closes on the local symmetries:
\ba
[ \delta_1 , \delta_2 ] = \nonumber & {\delta}_{gct}( \xi^{\mu} = -i ({\bar{\epsilon}}_1 \gamma^{\mu} \epsilon_2 )) \nonumber \\ 
+& \delta_{tens} (\Lambda^r_{\mu} = -\frac{1}{2} v^r \xi_{\mu} -\xi^{\nu} B^r_{\mu\nu}) \nonumber \\
+& \delta_{SO(n)}(A^{mn} = \xi^{\mu} x^{mr}(\partial_{\mu} x^n_r ) ) \nonumber \\
+& \delta_{Lorentz} (\Omega^{ab}  = -\xi_{\mu} (\omega^{\mu a b} - v_r H^{r \mu a b})) 
\label{susyalg}
\ea 
To this order, one can not  see the local supersymmetry transformation in the gauge algebra, since the expected parameter,  $\xi^\mu \Psi_\mu$, is generated  by bosonic variations. As usual, the spin connection satisfies its equation of motion, that to lowest order in the fermion  fields is 
\be 
D_\mu e_\nu{}^a - D_\nu e_\mu{}^a =0\quad 
\ee 
and implies the absence of torsion.

Completing these equations will require terms cubic in the fermion fields in the fermionic equations, and  terms quadratic in the fermion fields in their supersymmetry transformations. Supersymmetry will then determine corresponding modifications of the bosonic equations, and the (anti)self-duality conditions (\ref{selfdual}) will also be modified by terms quadratic in the fermion fields. Supercovariance actually fixes all terms containing the gravitino in the first-order equations and in the supersymmetry variations of fermion fields. The UV sensitivity of this scenario would be interesting to calculate for chernons (a future project). 

\subsection{Holographic duality}
\label{holodual}

In this subsection we review holographic duality to introduce gravity. This duality holds between AdS gravity and conformal field theory (CFT) in such a way that the 4D boundary CFT, whose definition and formal structure is well understood, could provide a fully nonperturbative definition of 5D quantum gravity. For example, the holographic dual of 6D LST corresponds to a 7D gravitational background with flat string-frame metric and the dilaton linear in the extra dimension \cite{aharony&al:1998}. 

The 5D heterotic string compactified on $K3\times S1$ should be equivalent to 11D supergravity compactified on a Calabi-Yau threefold \cite{anto&al:1996}. It yields the main properties of the holographic dual of 6D little string theory. Introducing back gravity weakly coupled, one has to compactify the extra dimension on an interval and place the SM on one of the boundaries, in analogy with the Randall-Sundrum model \cite{ran&sun:1999} on a slice of a 5D anti-de Sitter bulk.

In \cite{Antoniadis:2011qw, poko&al:2018} holographic dual models have been disclosed. The holographic dual of 6D LST can be approximated by a 5D model, in which the Lagrangian in the bulk takes the following form 
\be
e^{-1}\mathcal{L}_{LST} = - \widetilde{M}_5^3 \mathcal{R}-\frac{1}{3} (\partial_M \tilde{\Phi}) (\partial^M \tilde{\Phi}) - e^{\frac{2}{3} \frac{\tilde{\Phi}}{\widetilde{M}_5^{3/2}}} \Lambda
\ee
in the Einstein frame, where $e= \det(e_M^m)$, $\tilde{\Phi}$ is the dilaton and $\Lambda$ is a constant. Upon redefining
\be\label{correspondence}
\widetilde{\Phi}  = \sqrt{ \frac{3}{2}} \Phi \quad , \quad  \widetilde{M}_5^3 =  \frac{1}{2} M_5^3
\ee
and setting the gravitational coupling $\kappa$ in five dimensions equal to one ($\kappa^2=1/{M^3_5}$, where $M_5$ is the Planck mass in five dimensions), we obtain the Lagrangian for the canonically normalized dilaton $\Phi$
\be \label{lstpot}
e^{-1}\mathcal{L}_{LST} = - \frac{1}{2} \mathcal{R}-\frac{1}{2} (\partial_M \Phi) (\partial^M \Phi) - e^{\frac{2}{\sqrt{3}} \Phi} \Lambda\,
\ee
We thus observe that the potential that arises from LST is equal to the potential in 
\be \label{eff}
e^{-1} \mathcal{L}_{dilaton} = -\frac{1}{2} (\partial_M \Phi) (\partial^M \Phi) +3g^2A \Bigg( \frac{A}{4}e^{\frac{2}{\sqrt3}\Phi}+Be^{-\frac{1}{\sqrt3}\Phi} \Bigg)\,
\ee
for a scalar that belongs to a gauged $\mathcal{N}=2, D=5$ Maxwell multiplet coupled to supergravity, upon making the identification
\be \label{condit}
\frac{3}{4}g^2 A^2 = - \Lambda \quad , \quad B=0\,
\ee
We then have
\be \label{comp}
P_0 = A e^{\frac{1}{\sqrt{3}} \Phi} \quad , \quad P^a = - \frac{A}{2} e^{\frac{1}{\sqrt{3}} \Phi} \,
\ee

Moreover, it is known that the dilaton potential in (\ref{lstpot}) exhibits a runaway behavior and does not have a 5D maximally symmetric vacuum, but has a 4D Poincar\'e vacuum in the linear dilaton background
\be\label{dilaton}
\Phi = C y 
\ee
where $y>0$ is the fifth dimension and $C$ a constant parameter. The background bulk metric is then
\be\label{metric}
ds^2 =e^{-\frac{2}{\sqrt{3}}Cy} (\eta_{\mu \nu}dx^\mu dx^\nu+dy^2)
\ee
where $\eta_{\mu \nu}$ is the Minkowski metric of 4D space, under the fine-tuning condition 
\be \label{fine}
C=\frac{gA}{\sqrt{2}}
\ee 

The final Lagrangian then takes the form \cite{poko&al:2018}
\be\label{lagrangian}
\begin{array}{rcl}
	e^{-1} \tilde{\mathcal{L}} &=& - \frac{1}{2} \mathcal{R}(\omega) -\frac{1}{2} (\partial_M \Phi) (\partial^M \Phi)  - \frac{1}{8} e^{\frac{4}{\sqrt{3}}\Phi} F_{MN}^0 F^{MN 0} - \frac{1}{4}  e^{-\frac{2}{\sqrt{3}}\Phi} F_{MN}^1 F^{MN 1}
	\crbig
	& &  - \frac{1}{2} \bar{\psi}_M^i \Gamma^{MNP}\mathcal{D}_N \psi_{Pi} - \frac{1}{2} \bar{\lambda}^{i}\tilde{\slashed {\mathcal{D}}}  \lambda_i- \frac{i}{2} (\partial_N\Phi) \, \bar{\lambda}^{i} \Gamma^M \Gamma^N \psi_{Mi}
	\crbig
	& &  - \frac{\upsilon^0}{16\sqrt{2}}\, e^{\frac{5}{\sqrt{3}}\Phi}\, \bar{\lambda}^{i} \Gamma^M \Gamma^{\Lambda P} \psi_{M i} \, F_{\Lambda P}^0 - \frac{\upsilon^1}{8\sqrt{2}}   e^{-\frac{1}{\sqrt{3}}\Phi} \bar{\lambda}^{i} \Gamma^M \Gamma^{\Lambda P} \psi_{M i} \, F_{\Lambda P}^1 
	\crbig
	& &  - \frac{i\upsilon^0}{64}\sqrt{\frac{2}{3}}e^{\frac{5}{\sqrt{3}}\Phi}\,  \bar{\lambda}^{i} \Gamma^{MN} \lambda_i \, F_{MN}^0 - \frac{i\upsilon^1}{32}\sqrt{\frac{2}{3}}  e^{-\frac{1}{\sqrt{3}}\Phi}   \bar{\lambda}^{i} \Gamma^{MN} \lambda_i \,  F_{MN}^1 
	\crbig
	& & + \frac{1}{6 \sqrt 6}\, e^{\frac{5}{\sqrt{3}}\Phi}\,  C_{IJK}\epsilon^{MNP\Sigma \Lambda} F_{MN}^I F_{P \Sigma}^J A_\Lambda^K 
	\crbig
	& &  - \frac{3i\upsilon^0}{32\sqrt{6}}e^{\frac{5}{\sqrt{3}}\Phi} \,  [\bar{\psi}_M^i \Gamma^{MNP\Sigma} \psi_{Ni}F_{P\Sigma}^0+2 \bar{\psi}^{Mi}\psi_i^N F_{MN}^0]
	\crbig
	& &  - \frac{3i\upsilon^1 }{16\sqrt{6}}   e^{-\frac{1}{\sqrt{3}}\Phi}  [\bar{\psi}_M^i \Gamma^{MNP\Sigma} \psi_{Ni}F_{P\Sigma}^1+2 \bar{\psi}^{Mi}\psi_i^N F_{MN}^1]
	\crbig
	& &  + \frac{3g^2}{4}e^{\frac{2}{\sqrt3}\Phi}  - \frac{ig\sqrt{6}}{8}   \, e^{\frac{1}{\sqrt3}\Phi}   \, \bar{\psi}^i_M \Gamma^{MN} \psi_N^j \delta_{ij}
	\crbig
	&& + \frac{g}{2\sqrt{2}} \,  e^{\frac{1}{\sqrt3}\Phi}   \, \bar{\lambda}^{i} \Gamma^M \psi_M^j \delta_{ij} - \frac{ig}{4\sqrt{6}} \, e^{\frac{1}{\sqrt3}\Phi}   \, \bar{\lambda}^{i} \lambda^{j} \delta_{ij} 
	\crbig
	&& 
	+ \textrm{\,(4--fermion terms)}
\end{array}
\ee
This Lagrangian has three free parameters: $g$, $\upsilon^0$ and $\upsilon^1$.

The spectrum of the above model can be decomposed using the 4D Poincar\'e invariance of the linear dilaton vacuum solution and should form obviously $\mathcal{N}=1$ supermultiplets. It is known that every 5D field should give rise to a 4D zero mode and a continuum starting from a mass gap fixed by the linear dilaton coefficient $C=g/\sqrt{2}$. Using the results of \cite{Antoniadis:2011qw} and the correspondence (\ref{correspondence}), one finds that the parameter $\alpha$ of~\cite{Antoniadis:2011qw} is given by $\alpha=\sqrt{3}C$ and that the mass gap $M_{\rm gap}$ is 
\be
M_{\rm gap}=\frac{\sqrt{3}}{2\sqrt{2}}g
\ee
The continuum becomes an ordinary discrete Kaluza-Klein (KK) spectrum on top of the mass gap, when the fifth coordinate $y$ is compactified on an interval \cite{Antoniadis:2011qw}, allowing to introduce the Standard Model (SM) on one of the boundaries. This spectrum is valid for the graviton, dilaton and their superpartners by supersymmetry. Notice that the 5D graviton zero-mode has five polarisations that correspond to the 4d graviton, a KK vector and the radion. For the rest of the fields, special attention is needed because of the gauging that breaks half of the supersymmetry around the linear dilaton solution.

Indeed, one of the 4D gravitini acquires a mass fixed by $g$, giving rise to a massive spin 3/2 multiplet together with two spin 1 vectors. These are the 5D graviphoton and the additional 5D vector that have non-canonical, dilaton dependent, kinetic terms, as one can see from the Lagrangian~(\ref{lagrangian}). Using the background (\ref{dilaton}), (\ref{metric}), one finds that the $y$-dependence of the vector kinetic terms at the end of the first line of~(\ref{lagrangian}) is $\exp{\{\pm\sqrt{3}C\}}$ with the plus (minus) sign corresponding to the 5D graviphoton $I=0$ (extra vector $I=1$). It follows that they both acquire a mass given by the mass gap. The graviton multiplet is shown in table \ref{tab:table6}.
\begin{table}[H]
	\begin{center}
		\captionsetup{width=.8\linewidth}
		\begin{tabular}{|l|l|} 
			\hline
			Multiplet & Particle, Sparticle \\ 
			\hline
			gravitonmultiplet & $g_{\mu\nu}$, gravitino; graviphoton, spin 1 vector; dilaton	\\
			\hline
		\end{tabular}
		\caption{\small The gravitonultiplet.} 
		\label{tab:table6}
	\end{center}
\end{table}
We conclude with some comments on some possible phenomenological implications of the above lagrangian. One has to dimensionally reduce it from $D=5$ to $D=4$, upon compactification of the $y$-coordinate. Moreover, one has to introduce the SM, possibly on one of the boundaries, a radion stabilization mechanism and the breaking of the leftover supersymmetry. An interesting possibility is to combine all of them along the lines of the stabilisation proposal of~\cite{Cox:2012ee} based on boundary conditions.

There are several possibilities for Dark Matter (DM) candidates in this gravitational sector. There are two gravitini that, upon supersymmetry breaking can recombine to form a Dirac gravitino~\cite{Benakli:2014daa} or remain two different Majorana ones. Depending on the nature of their mass, the exact freeze-out mechanism will be different. There are three possible dark photons $A^0_\mu$, $A^1_\mu$ and the KK $U(1)$ coming from the 5D metric that could also be DM or their associated gaugini could also play a similar role, again depending on the compactification of the extra coordinate, on how supersymmetry breaking is implemented, as well as on the radion stabilisation mechanism. In general there could be a very rich phenomenology in the gravitational sector.

\section{Conclusions}
\label{conclusions}

The main results of this little string scenario are \\

(1) new level of topological matter between $\Lambda_{cr}$ and string scale $M_{string}$ with 3D Chern-Simons interaction,  \\

(2) Wess-Zumino supersymmetric Lagrangian is extended to charged and colored chernons in (\ref{chiralcharge}) and (\ref{chiralcolor}) and to include CS chernon binding interaction (\ref{Vmcs}). The octet of gluons emerges from fractionally charged chernon-antichernon pairs as indicated in (\ref{gluons}). First generation matter and the dark sector are formed of chernons by CS interaction binding. The standard model is derived heuristically. Our implementation of unbroken supersymmetry is wished to reward in the calculations (in a new project). A new bosonic QCD spectroscopy is predicted by the gluon triplet states in table \ref{tab:table1}. (These particles may have been found under a different classification name),  \\

(3) supergravity is obtained by local supersymmetry, see table \ref{tab:table5}. Holography connects SUGRA and LST, see subsection \ref{holodual} and table \ref{tab:table6}.  \\

Earlier results of this scenario include (i) supergravity based inflation in the early universe followed by a quantum statistical mechanism for baryon asymmetry with $\frac{n_B}{n_{\gamma}} \gg 1$ \cite{rai_03, rai_03a}, (ii) economic unification of matter and interactions based on few supermultiplets, and (iii) a brief discussion on string and preon symmetries in \cite{rai_06}.

In the absence of MSSM superpartners the present model is a noteworthy, prediction verifiable candidate for BSM physics.

\newpage
\appendix
\section{The basic idea of the model}
\label{basicidea}

\begin{figure}[H]
	\captionsetup{width=0.8\textwidth}
	\centering
	\includegraphics[scale=0.3]{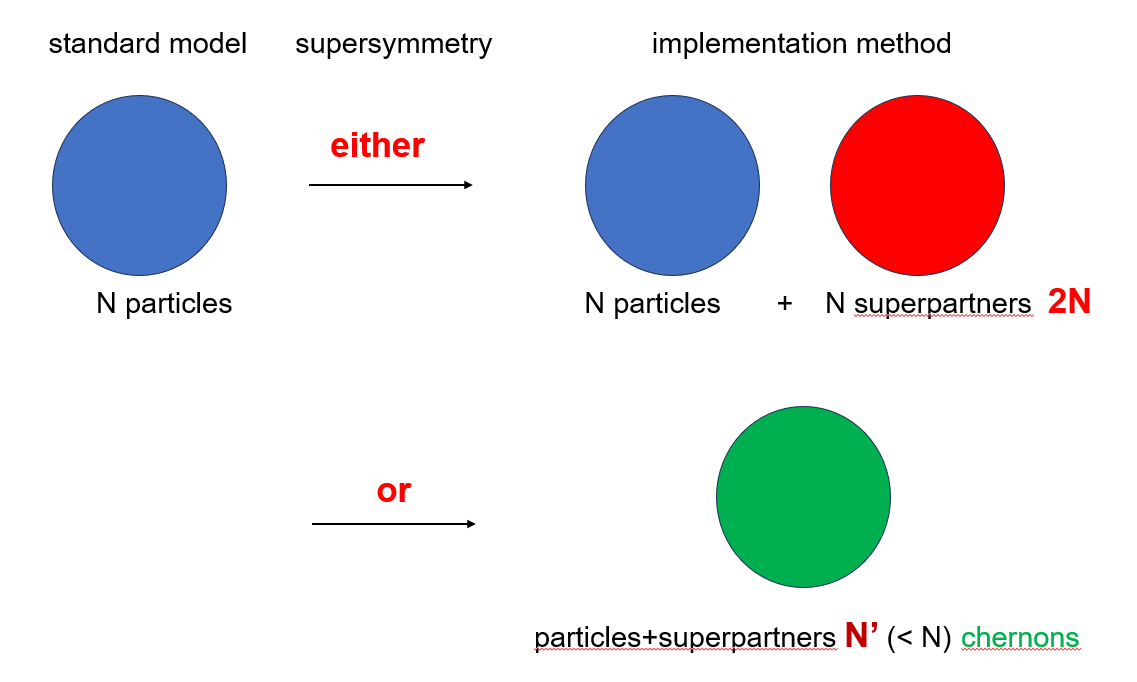}
	\caption{\small On the top/right is the standard minimal supersymmetric model logical structure with the number of SM particles doubled by introducing the superpartners, disclosing a risk for "double counting". Furthermore, none of the SM superpartners (red) have been observed in nature so far. Bottom picture (green) indicates that by splitting quarks and leptons into three $m$ fermions plus some bosons in table \ref{tab:table1} the composite scenario has fewer elementary particles organized in supermultiplets.}
	\label{fig:figurea}
\end{figure}

We attempt to visualize the basic idea behind the model in figure \ref{fig:figurea}. 

\vskip 1cm

\begin{thebibliography}{99}

\bibitem{rai_00} Risto Raitio, A Model of Lepton and Quark Structure. Physica Scripta, 22, 197 (1980). \href{https://dx.doi.org/10.1088/0031-8949/22/3/002}{\tt PS22, 197}, \href{http://vixra.org/abs/1903.0224}{viXra:1903.0224}
\footnote{~The model was conceived in November 1974 at SLAC. I proposed that the c-quark would be a gravitational excitation of the u-quark, both composites of three 'subquarks'. The idea was opposed by the community and was therefore not written down until five years later.}

\bibitem{rai_01} Risto Raitio, Supersymmetric preons and the standard model, Nuclear Physics B931 (2018) 283–290. \href{https://doi.org://10.1016/j.nuclphysb.2018.04.021}{\tt doi:10.1016/j.nuclphysb.2018.04.021} \href{https://arxiv.org/pdf/1805.03013.pdf}{\tt arXiv:1805.03013} 

\bibitem{Wess_Z} J. Wess and B. Zumino, Supergauge transformations in four dimensions, Nucl. Phys. B 70 (1974) 39. \href{https://doi.org/10.1016/0550-3213(74)90355-1}{\tt 10.1016/0550-3213(74)90355-1}

\bibitem{Peccei_Q} Roberto D. Peccei and Helen R. Quinn,  CP Conservation in the Presence of Pseudoparticles, Phys. Rev. Lett. 38 (25) 1440–1443 (1977). 

\bibitem{Papa_A_F} J. Papavassiliou, A. C. Aguilar, and M. N. Ferreira, Theory and phenomenology of the three-gluon vertex, Talk presented by J.P. at the 19th International Conference on Hadron Spectroscopy and Structure (HADRON 2021), 26 July- 1 August 2021, Mexico City, Mexico.
\href{https://arxiv.org/pdf/2201.08496.pdf}{\tt arXiv:2201.08496}

\bibitem{rai_05} Risto Raitio, The fate of supersymmetry in topological quantum field theories.
\href{https://arxiv.org/pdf/2307.13017.pdf}{\tt arXiv:2307.13017}

\bibitem{rai_03} Risto Raitio, A Scenario for Asymmetric Genesis of Matter, Journal of High Energy Physics, Gravitation and Cosmology, 9, 654-665 (2023). 
\href{https://doi.org/10.4236/jhepgc.2023.93053}{\tt 10.10.4236/jhepgc.2023.93053}

\bibitem{rai_03a} Risto Raitio, A Chern-Simons model for baryon asymmetry, Nuclear Physics B Volume 990, May 2023, 116174. \href{https://arxiv.org/pdf/2301.10452.pdf}{\tt arXiv:2301.10452}

\bibitem{Beli_D_F_H} H. Belich, O. M. Del Cima, M. M. Ferreira Jr. and J. A. Helay\"{e}l-Neto, Electron-Electron Bound States in Maxwell-Chern-Simons-Proca QED$_3$, Eur. Phys. J. B 32, 145–155 (2003).
doi:https://doi.org/10.1140/epjb/e2003-00083-9
\href{https://arxiv.org/pdf/hep-th/0212285.pdf}{\tt arXiv:hep-th/0212285}

\bibitem{Kogan} Ya. I. Kogan, Bound states of fermions and superconducting ground state in a 2+1 gauge theory with a topological mass term, JETP Lett. 49, 225 (1989).

\bibitem{Dobroliubov} M. I. Dobroliubov, D. Eliezer, I. I. Kogan, G.W. Semenoff and R.J. Szabo, The Spectrum of Topologically Massive Quantum Electrodynamics, Mod. Phys. Lett. A, 8, 2177 (1993).
doi:https://doi.org/10.1142/S0217732393001902

\bibitem{LMS} A. Losev, G. Moore and S.L. Shatashvili, M \& m’s, Nucl. Phys. B522 (1998) 105. 
doi:https://doi.org/10.1016/S0550-3213(98)00262-4 ~ 
\href{https://arxiv.org/pdf/hep-th/9707250.pdf}{\tt arXiv:hep-th/9707250}

\bibitem{deser.stelle.kay:1977} S. Deser, J. H. Kay and K. S. Stelle, Renormalizability Properties of Supergravity, Phys. Rev. Lett. 38, 527 (1977).
doi:https://doi.org/10.1103/PhysRevLett.38.527

\bibitem{ricc.sagn:1998} Fabio Riccioni1 and Augusto Sagnotti, Some Properties of Tensor Multiplets in Six-Dimensional Supergravity, Nucl.Phys.Proc.Suppl. 67 (1998) 68-73.
doi:https://doi.org/10.1016/S0920-5632

\bibitem{Vafaetal.2016} Lakshya Bhardwaj, Michele Del Zotto, Jonathan J. Heckman, David R. Morrison, Tom Rudelius and and Cumrun Vafa, F-theory and the Classification of Little Strings, Phys. Rev. D 93, 086002 (2016).
\href{https://doi.org/10.1103/PhysRevD.93.086002}{\tt doi:10.1103/PhysRevD.93.086002}.
\href{https://arxiv.org/pdf/1511.05565v3.pdf}{\tt arXiv:1511.05565v3}

\bibitem{DelZottoetal.2022} Michele Del Zotto, Muyang Liu, and Paul-Konstantin Oehlmann, 6D Heterotic Little String Theories and F-theory Geometry: An Introduction, Contribution to the proceedings of String Math 2022, University of Warsaw, Poland, July 11-15, 2022.
\href{https://arxiv.org/pdf/2303.13502.pdf}{\tt arXiv:2303.13502}

\bibitem{ahmedaetal.2023} Hamza Ahmeda, Paul-Konstantin Oehlmanna and Fabian Ruehle, T-Duality and Flavor Symmetries in Little String Theories. \href{https://arxiv.org/pdf/2311.02168.pdf}{\tt arXiv:2311.02168}

\bibitem{aharony&al:2004} O. Aharony, A. Giveon and D. Kutasov, LSZ in LST, Nucl. Phys. B 691, 3 (2004) and references therein.
\href{https://arxiv.org/pdf/hep-th/0404016.pdf}{\tt arXiv:hep-th/0404016}

\bibitem{Vafa.1996} Cumrun Vafa, Evidence for F-theory Nucl. Phys.B 469, 403-418 (1996).
\href{https://doi.org/10.1016/0550-3213(96)00172-1}{doi:10.1016/0550-3213(96)00172-1}. 
\href{https://arxiv.org/pdf/hep-th/9602022.pdf}{\tt arXiv:hep-th/9602022}

\bibitem{Argyresetal.2022} Philip C. Argyres, Mario Martone, Michael Ray, Dirac pairings, one-form symmetries and Seiberg-Witten geometries,
\href{https://doi.org/10.1007/JHEP09%282022%29020}{\tt doi:10.1007/JHEP09\%282022\%29020}.
\href{https://arxiv.org/pdf/2204.09682v3.pdf}{\tt arXiv:2204.09682v3}

\bibitem{seib-tayl.2011} Nathan Seiberg and Washington Taylor, Charge lattices and consistency of 6D supergravity,  J. High Energ. Phys. 2011, 1 (2011). \href{https://doi.org/10.1007/JHEP06(2011)001}{\tt 10.1007/JHEP06(2011)001}.
\href{https://arxiv.org/pdf/1103.0019.pdf}{\tt arXiv:1103.0019}

\bibitem{Bhardwaj:2015xxa} Lakshya Bhardwaj, Classification of 6d N = (1,0) gauge theories. 
doi:\href{http://dx.doi.org/10.1007/JHEP11(2015)002}{\tt JHEP 11 (2015) 002}.
\href{http://arxiv.org/abs/1502.06594}{\tt arXiv:1502.06594}

\bibitem{Seiberg:1996qx} Nathan Seiberg, Nontrivial fixed points of the renormalization group in six-dimensions, Phys. Lett. B 390 (1997) 169-171.
doi:\href{https://doi.org/10.1016/S0370-2693(96)01424-4}{\tt 10.1016/S0370-2693(96)01424-4}~
\href{http://arxiv.org/abs/hep-th/9609161}{\tt arXiv:hep-th/9609161}

\bibitem{romans} L.J. Romans, Self-duality for interacting fields: Covariant field equations for six-dimensional chiral supergravities, Nucl. Phys. B276 (1986) 71. doi:https://doi.org/10.1016/0550-3213(86)90016-7

\bibitem{aharony&al:1998} O. Aharony, M. Berkooz, D. Kutasov and N. Seiberg, Linear dilatons, NS5-branes and holography, JHEP 9810, 004 (1998). [arXiv:hep-th/9808149].

\bibitem{anto&al:1996} I. Antoniadis, S. Ferrara, T.R. Taylor, N=2 Heterotic Superstring and its Dual Theory in Five Dimensions, Nucl. Phys.B 460, 489-505 (1996). 
doi:https://doi.org/10.1016/0550-3213(95)00659-1 
\href{http://arxiv.org/abs/hep-th/9511108}{\tt arXiv:hep-th/9511108}

\bibitem{ran&sun:1999} L. Randall and R. Sundrum, “A Large mass hierarchy from a small extra dimension,” Phys. Rev. Lett. 83 (1999) 3370. doi:10.1103/PhysRevLett.83.3370
\href{http://arxiv.org/abs/hep-ph/9905221}{\tt arXiv:hep-ph/9905221}

\bibitem{Antoniadis:2011qw} I.~Antoniadis, A.~Arvanitaki, S.~Dimopoulos and A.~Giveon, Phenomenology of TeV Little String Theory from Holography,
Phys. Rev. Lett. 108 (2012) 081602.
doi:10.1103/PhysRevLett.108.081602 \\
\href{http://arxiv.org/abs/1102.4043}{\tt arXiv:/1102.4043}

\bibitem{poko&al:2018} Ignatios Antoniadis, Antonio Delgado, Chrysoula Markou and Stefan Pokorski, The effective supergravity of little string theory, Eur. Phys. J. C (2018) 78:146.
doi:https://doi.org/10.1140/epjc/s10052-018-5632-4

\bibitem{Cox:2012ee} P.~Cox and T.~Gherghetta, Radion Dynamics and Phenomenology in the Linear Dilaton Model, JHEP 1205 (2012) 149.
doi:10.1007/JHEP05(2012)149
arXiv:hep-ph/1203.5870 

\bibitem{Benakli:2014daa} K.~Benakli, (Pseudo)goldstinos, SUSY fluids, Dirac gravitino and gauginos, EPJ Web Conf. 71 (2014) 00012.
doi:10.1051/epjconf/20147100012
\href{https://arxiv.org/pdf/1402.4286.pdf}{\tt arXiv:1402.4286}

\bibitem{rai_06} Risto Raitio, Stringy phenomenology with preon models, Journal of High Energy Physics, Gravitation and Cosmology Vol.10 No.1, December 29, 2023.
doi: 10.4236/jhepgc.2024.101001
\href{https://arxiv.org/pdf/2310.01463.pdf}{\tt arXiv:2310.01463} 

\end{thebibliography}
\end{document}